\documentclass[letterpaper,12pt]{article}
\usepackage{tabularx} 
\usepackage{amsmath}  
\usepackage{float}
\usepackage{indentfirst}
\usepackage{graphicx} 
\usepackage[margin=1in,letterpaper]{geometry} 
\usepackage{caption}
\usepackage{color,soul}
\usepackage{multicol}
\usepackage{amsfonts}
\usepackage{amssymb}
\usepackage{authblk}
\usepackage{graphicx}
\usepackage{cite} 
\usepackage[final]{hyperref} 
\usepackage{tensor}
\hypersetup{
	colorlinks=true,       
	linkcolor=red,        
	citecolor=blue,        
	filecolor=magenta,     
	urlcolor=blue
}
\date{}
\begin{document}
\title{Entropy variation of rotating BTZ black hole under Hawking radiation}
\author[*,1,2]{Shad Ali}
\author[1]{Muhammad Arshad Kamran}
\author[3]{Misbah Ullah Khan}
\small{
\affil[1]{Department of Physics, University of Okara, 56300, Punjab, Pakistan.}
\affil[2]{Department of Astronomy, Xiamen University, Xiamen, Fujian 361005, China}
\affil[3]{Department of Center for Nanosciences,University of Okara,56300, punjab, pakistan.}
}
\maketitle
\begin{abstract}
An axially symmetric rotating Banados-Teitelboim-Zanelli (BTZ) black hole is considered for comprehending its interior information. The largest space-like hyper-surface is chosen to estimate its maximal interior volume as a time-dependent quantity. Similarly, the quantum mode entropy of the scalar field associated with this volume is found to increase with Eddington time. An evolution relation between the variation of quantum mode entropy and Bekenstein-Hawking entropy is obtained for an infinitesimal time interval. On comparing to higher-dimensional black holes, the characteristic feature of this evolution relation is its divergent character.

\textbf{Keywords:} BTZ black hole, interior volume and entropy, information paradox, evaporation.

\end{abstract}

\vfill {\footnotesize ~\\{*}Corresponding Author\\ E-mail: shad.ali88@yahoo.com}
\newpage

\section{Introduction}
 In Minkowski space-time, the volume of an n$-$dimensional sphere $S^n$ can be viewed as the volume of an (n + 1)$-$dimensional spacelike surface $\sum$ bounded by $S^n$. For example, the volume inside a two-sphere $S^2$ is the volume of the largest $3-$dimensional spacelike spherically-symmetric surface $\sum$ bounded by $S^2$. Its maximal interior volume is expressed as $V=\frac{3}{4}\pi R^3$. It means that the maximal interior volume bounded by the largest $3-$dimensional space-like surface $\sum$ is $V$. A suitable space-like hyper-surface enclosing the maximal interior volume can be identified by two equivalent definitions \cite{Christodoulou:2014yia}:
 
\begin{itemize}
\item This $\sum$ must be the largest spherically symmetric surface (If there exist other space-like surfaces bounded by $S^2$) and
\item This $\sum$ must lie on the same simultaneity surface as $S^2$.
\end{itemize}

In curved space-time, the investigations of interior volume is constrained by two problems: First, the  simultaneity condition is not valid due to lacking of space-time flatness. Second, the interchange of  the space-times coordinates in the interior of black hole horizon. So, one can't choose a largest space-like surface to that bounds the maximal interior volume of black hole. There are several proposals for investigation of  black hole interior volume e.g. the interior volume of a stationary black hole as slicing invariant was found by Parikh \cite{Parikh:2005qs}. The thermodynamic volume of a black hole is elaborated as conjugate to cosmological constant in Refs. \cite{Kastor:2009wy, Kubiznak:2012wp, Gunasekaran:2012dq}, the volume for dynamical black hole in terms of Kodama vector is proposed in Refs. \cite{Hayward:1997jp, Hayward:1999ek} and the vector volume is discussed in Ref. \cite{Ballik:2013uia}.

Christodoulou and Rovelli (CR) \cite{Christodoulou:2014yia, Christodoulou:2016tuua} pioneered to calculate the largest volume of space-like spherically symmetric hyper-surface bounded by a Schwarzschild black hole. Considering Schwarzschild black hole with static exterior and dynamical interior they obtained a time-dependent interior volume as
\begin{equation}\label{eq1}
V_{\sum}=3\sqrt{3}\pi m^2 v,
\end{equation}
here $v$ is the Eddington time. The maximal volume given in Eq. (\ref{eq1}) is termed as CR volume that increases with Eddington time. This work is also extended to other types of black holes in Refs. \cite{Bhaumik:2016sav, Yang:2018arj, Ong:2015tua, Ong:2015dja, Ali:2018sqk, Wang:2018dvo, Bengtsson:2015zda, Zhang:2019pzd, Ali:2020olc}.

Based on the CR's volume, Baocheng Zhang \cite{Zhang:2015gda} considered interior quantum modes of scalar field and investigated the entropy inside the Schwarzschild black hole as
\begin{equation}\label{entropy}
S_{\sum}=\frac{\pi^2}{45}T^3 V_{\sum},
\end{equation}
here $T$ is the horizon temperature. Majhi and Samanta \cite{Majhi:2017tab} followed constraint analysis to investigate black hole entropy but their investigations were based on an ansatz metric rather than actual metric. The main difference between the two methods is discussed in Ref. \cite{Zhang:2017aqf}. As our main purpose is to study the interior information of rotating BTZ black hole so, here we will not probe the difference of two articles.

Eq. (\ref{entropy}) shows the quantum mode entropy is proportional to interior volume of black hole and hence to Eddington time. This relation of Eddington time with interior volume and quantum mode entropy has special role in affecting the statistical quantities inside the black hole. It is a good point for probing the information from the interior of black hole. The investigation of  a relation between the interior and exterior entropy of black hole could led to discuss the black hole evaporation \cite{Parikh:2005qs}. So, using the notion of interior entropy with two important assumptions of \textit{"black hole radiations as black body radiations"} and \textit{"the emission rate as quasi-static"} it was found that the rate of change of interior entropy is related to the Bekenstein-Hawking entropy by a simple relation as discussed in Refs. \cite{Wang:2018dvo,  Ali:2018sqk, Ali:2019icq}. 

In Refs. \cite{Zhang:2019pzd, Ali:2020olc}, the interior volume of BTZ black hole is found to increase with $v$. It signifies the possible extension of our work to $(2+1)-$ dimensional BTZ black hole solutions. So, after obtaining the interior volume and quantum modes entropy, an evolution relation is calculated between the two types of entropy by using the two assumptions. This evolution relation is a linear function of black hole mass and comparing with higher-dimensional black holes it showed a different notion for understanding the black hole evaporation and its interior information.

In this work, we followed the above discussion to probe the interior information of rotating BTZ black hole. The structure of this paper is such that in the next section (\ref{Sec.2}), we discussed the metric and interior volume for rotating BTZ black hole. In section (\ref{Sec.3}), we discussed the entropy in the interior of the rotating BTZ black hole and its evolution relation with Bekenstein-Hawking entropy. Remarks and discussion are  in section (\ref{Sec.4}) and the conclusions are added in section (\ref{Sec.5}).

\section{Metric of rotating BTZ black hole and Interior volume}\label{Sec.2}

BTZ black hole is a solution of Einstein-Maxwell equations in $(2+1)-$dimensional space-time with negative cosmological constant and constant negative curvature \cite{Banados:1992wn, Banados:1992gq, Birmingham:2001dt}. According to Carlip \cite{Carlip:1995qv}, the classical and quantum properties of $(2+1)-$dimensional BTZ black hole shares many characteristics to $(3+1)-$ dimensional black holes. The metric of $(2+1)-$dimensional rotating BTZ black hole is defined as 
\begin{equation}\label{metric}
ds^2=-f(r) dt^2+\frac{dr^2}{f(r)}+r^2 (N^\phi (r) dt+d\phi)^2,
\end{equation}
where
\begin{equation}
f(r)=-m+\frac{r^2}{l^2}+\frac{J^2}{4 r^2}, \quad N^\phi (r)= -\frac{J}{2r^2},  \left (\left | J \right |\leq ml  \right ),
\end{equation}  
here $l$ is the AdS radius related to cosmological constant $\Lambda=-\frac{1}{l^2}$, $\phi$ is the period (with range of $0\leq\phi\leq 2\pi$ for the representation of black hole space-time), $J$ is the azimuthal angular momentum corresponding to angular velocity $\Omega (r)$ and $m$ is the AdS mass of black hole. Since the metric of rotating BTZ black holes has an azimuthal symmetry so, the angular momentum is conserved under a coordinate transformation. The vanishing value of lapse function gives the locations of inner and outer horizons
\begin{equation}\label{radius}
r_\pm =\sqrt{\frac{l^2 m}{2}  \left(1\pm X\right)}, \qquad with \quad  X=\sqrt{1-\left(\frac{J}{l m}\right)^2},
\end{equation}
This Eq. (\ref{radius}) gives a BTZ black hole solution $r_{\pm}= l\sqrt{\frac{m}{2}}$ for  $X=0 \Rightarrow \frac{J}{l m}=1$. The mass $m$ and Bekenstein-Hawking entropy $S_{BH}$ of  rotating BTZ black hole are
\begin{equation}\label{Hawent}
m= \frac{r^2}{l^2}+\frac{J^2}{4 r^2}, \qquad S_{BH}=4\pi r_+,
\end{equation}
Using the lapse function, one can easily define the horizon temperature as
\begin{equation}\label{Temp}
T=\frac{f'(r_+)}{4\pi}=\frac{mX}{\pi\sqrt{2l^2m(1+X)}},
\end{equation}
here the dash $(')$ represents the derivative with respect to $r$. The angular velocity around the axis of rotation is
\begin{equation}\label{angularvelocity}
d\phi =\Omega (r) dt=  \frac{J}{2r_+ ^2}dt=-N^{\phi} (r)dt,
\end{equation}
Remember that this definition of angular velocity is valid only in case of $(2+1)-$dimensional space-time. This point enables us to calculate the number of quantum modes in the interior of the proposed scalar field. The Penrose and Misner super-radiation processes are active in the region of  Ergo-sphere with range $r_+<r<r_s$. The Heat capacities due to constant angular momentum J and constant angular velocity $C_\Omega$ are
\begin{equation}
C_J =T\left(\frac{\partial S_{BH}}{\partial T}\right)_J=C_o \frac{X}{2-X}\sqrt{\frac{1+X}{2}}, \quad C_\Omega =T\left(\frac{\partial S_{BH}}{\partial T}\right)_\Omega=4\pi\sqrt{\frac{m(1+X)}{2}},
\end{equation}
here  $C_o =4\pi r_+$ is the heat capacity at $J=0$. As $X<1$ so, for BTZ black hole the heat capacity is always positive. From above Eq. (\ref{metric}), we can write it as \cite{Carlip:1995qv}
\begin{equation}\label{BTZmet}
ds^2=-f(r) dv^2+2dvdr+r^2 (N^\phi (r) dt+d\phi)^2,
\end{equation}
As already stated that the shift function $N^\phi (r)$ is related to angular velocity hence. So, it will have a direct effect on $\Omega (r)$ and one can evolve the maximum or minimum value of red-shift related to clock-wise or counter clock-wise angular velocity. The metric Eq. (\ref{BTZmet}) satisfies the ordinary field equation and shows a uniformly rotating BTZ black hole geometry. A $2-$dimensional symmetric hyper-surface $\sum$ is the direct product of 1-sphere and a curve $\gamma$ in the $v-r$ plane \cite{Christodoulou:2014yia, Zhang:2019pzd}, i.e.
\begin{equation}
\sum=\gamma \times S^1  , \qquad \gamma \mapsto [v(\lambda), r(\lambda)]
\end{equation}
Here $\lambda$  is an arbitrary parameter whose initial and final values related to endpoints are
$$\lambda_i = [v(\lambda), r(\lambda)]_{\lambda=0}=(v_i, r_+), $$ $$ \lambda_f = [v(\lambda), r(\lambda)]_{\lambda=\lambda_f}=(v, 0),$$The Eddington-Finkelstein coordinates for BTZ black hole geometry from Eq. (\ref{BTZmet}) are
\begin{equation}\label{Eddmet}
ds^2=(-f(r){\dot{v}}^2+2 {\dot{v}} {\dot{r}}) d\lambda^2+r^2 (N^\phi (r) dt+d\phi)^2,
\end{equation}
The dot $(.)$ represents the differentiation concerning $\lambda$. A special case, that could lead this metric to a possible solution for interior volume of rotating BTZ black hole is to consider Eq. (\ref{angularvelocity}) with Eq. (\ref{Eddmet}).  As we considered an axially symmetric BTZ black hole with conserved angular momentum $J$ so, any deviation from axial symmetry will have negligible effects at the black hole horizon. Moreover, these forces will be more weaker in the interior of the black hole at $r=r_v$. Generally, the interior volume of rotating BTZ black hole can be defined as
\begin{equation}
V_{\sum}=\int ^{2\pi}_0 d\phi \int ^{\lambda_f} _0 \sqrt{(-f(r){\dot{v}}^2+2 {\dot{v}} {\dot{r}})}d\lambda,
\end{equation}
This equation can be maximized by $\lambda_f$ for a very large Eddington time so that the geodesic could spend maximum time at radius $r$. So, we can approximate the geodesic to initial and final transients with intermediate steady phase i.e. $\dot{r}=0$. Thus the maximum interior volume bounded by the largest hyper-surface of the rotating BTZ black hole is
\begin{equation}\label{Gvoleq}
V_{\sum} = 2{\pi} v\sqrt{-r_v ^2f(r_v)},
\end{equation}
The value of $r$ that maximizes the polynomial  $\sqrt{-r_v ^2f(r_v)}$ is referred to $r_v$ and the polynomial $\sqrt{-r_v ^2f(r_v)}$ called maximization factor for the interior volume of the black hole. This polynomial also guarantees non-negativity of the black hole interior volume under the radical sign. Its maximum value for other than Schwarzschild black hole can be evaluated by numerical method. From this equation, the interior volume of the black hole will be maximum for $r=r_v$ and we labeled it as maximal hyper-surface ranging from $0<r<r_+$. The maximal hyper-surface obtained by the maximization of polynomial $\sqrt{-r_v ^2f(r_v)}$ is
\begin{equation}\label{maxhyp}
r_v=\frac{\sqrt{l^2 m \left(\sqrt{3 X^2+1}+2\right)}}{\sqrt{6}},
\end{equation}
\begin{figure}
\begin{center}
\includegraphics[width=0.35\textwidth]{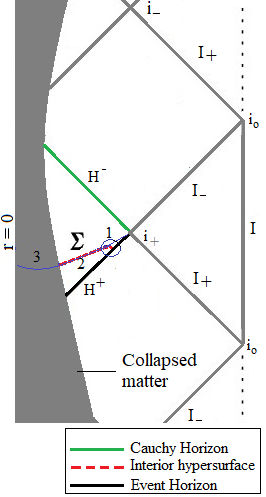}
\caption{Penrose diagram of rotating BTZ black hole formed under collapsed process.}
\label{image-1}
\end{center}
\end{figure}

\begin{figure}
\begin{center}
\includegraphics[width=0.7\textwidth]{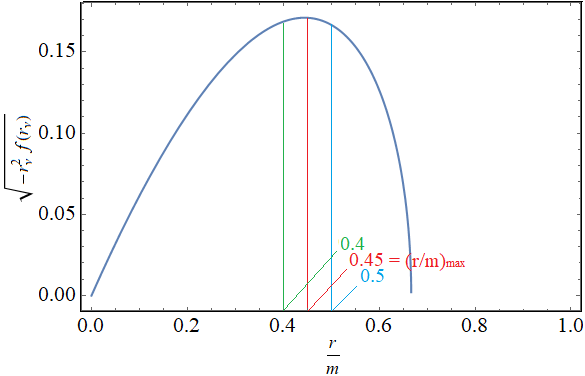}
\caption{Plot between maximization factor of volume $\sqrt{-r_v ^2f(r_v)}$  vs $\frac{r}{m}$. The position of maximal hyper-surface in the interior of black hole maximal volume reaches to $0.45m$ and \quad $J=0.5$.}
\label{image-2}
\end{center}
\end{figure}

The main idea for choosing the largest hyper-surface bounded by the $1-$sphere is sketched in Fig. (\ref{image-1}) using Penrose diagram. A space-like hyper-surface is drawn from event horizon to the center of collapsed matter and splitted in three sections as labeled by 1, 2 and 3. Section $(1)$ is the null part hyper-surface that connects the space-like hyper-surface to the event horizon of black hole. Section $(2)$ is a long stretch part of hyper-surface with nearly constant radius and section $(3)$ is the part of hypersurface that connects the hypersurface $r_v$ to the center of the collapsing object $r = 0$. By this terminology, section $(1)$ does not contribute to the interior volume of black hole and section $(3)$ is entirely inside the collapsed object. Since the space- time inside the collapsing object acquires a timelike killing vector field so, this part is a space-like hypersurface in this region and will have a finite contribution to the interior volume of the black hole. Hence, the total contribution of these two sections at large $v$ can be ignored while the volume associated with the section $(2)$ increases linearly with $v$. It means that we can consider section $(2)$ as largest space-like hyper-surface inside a rotating BTZ black hole denoted by $r_v$. Therefore, the volume bounded by black hole at $r=r_v$ is the maximum interior volume of a rotating BTZ black hole. In case of Schwarzschild black hole it is investigated as $r_v=1.5 m$ in Ref. \cite{Christodoulou:2014yia}.
Substituting this value of $r_v$ in Eq. (\ref{Gvoleq}), we get
\begin{equation}\label{BTZvoleq}
V_{\sum}=\frac{\pi v}{3} \sqrt{l^2 m^2 \left(-3 X^2+2 \sqrt{3 X^2+1}+7\right)-9 J^2},
\end{equation}
The position of the maximal hyper-surface is found at $0.45m$ as shown in Fig. (\ref{image-2}).

At large advance time, the largest spherically symmetric space-like hyper-surface bounds the volume given in Eq. (\ref{BTZvoleq}). It also shows that the maximal interior volume of the BTZ black hole is proportional to Eddington time i.e. BTZ black hole has a large scope of storing information in its interior. In Fig. (\ref{image-2}), the curve shows the maximum value of polynomial $\sqrt{-r_v ^2f(r_v)}$ for rotating BTZ black hole is at $\frac{r}{m}=0.45$ and all other points on the curve, $\sqrt{-r_v ^2f(r_v)}$ has less value of $\frac{r}{m}$ as compared to $0.45$. So, we considered the hyper-surface $\frac{r}{m}=0.45$ for the investigation of maximal interior volume. 

\section{Entropy of massless scalar field and its evolution relation with Bekenstein-Hawking Entropy} \label{Sec.3}
We have seen that the interior volume of a black hole is directly proportional to Eddington time. As there is a simple relation between the volume and entropy black hole. So, we approach forward to probe some new results related to the interior entropy of rotating BTZ black hole. The entropy of massless scalar field $\Phi$ in the interior volume bound by the maximal hyper-surface of BTZ black hole at $r=r_v$ can be found by using Wentzel-Kramer-Brillouin (WKB) approximation. We consider the massless scalar field as $\Phi=exp[-iET]exp [iI(\lambda, \phi)]$. So, expanding the Klein Gordon equation $\frac{1}{\sqrt{-g}}\partial_{\mu}(\sqrt{-g}g^{\mu v}\partial_v\Phi)=0$, we get
\begin{equation}
E^2-\frac{1}{(-f(r){\dot{v}}^2+2 {\dot{v}} {\dot{r}})}P^2 _{\lambda}-\frac{1}{r^2}P^2 _{\phi}=0,
\end{equation}
here $\partial_ \lambda I=P_{\lambda}$ and $\partial _ {\phi}I=P_{\phi} $ are used. This equation can be written as
\begin{equation}
P_{\lambda}=\sqrt{-f(r){\dot{v}}^2+2 {\dot{v}} {\dot{r}}}\sqrt{E^2-\frac{1}{r^2}P^2 _{\phi}},
\end{equation}
So, the total number of quantum states are 
\begin{equation}
g(E)=\frac{1}{(2\pi)^2}\int^{2\pi}_{0} d\phi \int d{\lambda} dp_{\lambda} dp_{\phi},
\end{equation}
\begin{equation}
=\frac{1}{(2\pi)^2}\int^{2\pi}_{0} d\phi \int d{\lambda}  \sqrt{-f(r){\dot{v}}^2+2 {\dot{v}} {\dot{r}}}\sqrt{E^2-\frac{1}{r^2}P^2 _{\phi}} dp_{\phi},
\end{equation}
\begin{equation}\label{Qmodes}
g(E)=\frac{E^2}{8\pi}\int^{2\pi}_{0} d\phi \int r  \sqrt{-f(r){\dot{v}}^2+2 {\dot{v}} {\dot{r}}}d{\lambda}=\frac{E^2}{8\pi}V_{\sum},
\end{equation}
we used the integral formula $\int_{0}^{a}{\sqrt{1-\frac{x^2}{a^2}}}dx=\frac{\pi}{4}a$ for calculating the total number of quantum states. The free energy for an inverse temperature $\beta=\frac{1}{T}$ is 
\begin{equation}\label{FE}
F(\beta)=\frac{1}{\beta}\int {ln(1-exp(-\beta E))}dg(E)=-\frac{\zeta (3)}{4\pi \beta^3}V_{\sum},
\end{equation}
here $\zeta(x)$ is Riemann zeta function. Finally, the entropy of massless scalar field as 
\begin{equation}\label{Ent1}
S_{\sum}=\beta ^2\frac{ \partial F(\beta)}{ \partial\beta}=\frac{3\zeta (3)}{2\beta^2}\sqrt{-r_v ^2f(r_v)}v,
\end{equation}
This Eq. (\ref{Ent1}) shows the entropy is also related linearly with $v$, it is a feature of this entropy to affect the statistical quantities in the interior of the BTZ black hole. To see this, let us consider two assumptions, it could lead us to investigate the evaluation relation between quantum modes entropy and Bekenstein-Hawking entropy. These assumptions are:

\begin{itemize}
\item {Black hole radiation as black body radiations: This assumption guarantees the black hole temperature seen from infinity as event horizon temperature so, one can use the Boltzmann law for the emission of radiation \cite{landsberg:1989}. In $(2+1)$ dimension space-time of rotating BTZ black hole, the Boltzmann law is 
\begin{equation}
\frac{dm}{dv}=-\sigma A T^3 \Rightarrow dv=-\frac{\beta^3 \gamma }{ A }dm,
\end{equation}here $A=\pi  l \sqrt{2 m (X+1)}$ is the area of event horizon and the $\beta$ value can be found from Eq. (\ref{Temp}).}
\item {The rate of radiation emission from black hole as quasi-static process i.e. $\frac{dm}{dv}<<1$. It means that the evaporation process is slow but Hawking temperature varies continuously. From this assumption, the thermal equilibrium between the scalar field and the event horizon of the black hole is preserved in an adiabatic process. It guarantees the differential form of radiation emission for an infinitely small interval of time.}
\end{itemize}

So, fixing these assumptions in Stefan Boltzmann law with the values of $\beta$ and $A$, we get the differential form of quantum mode entropy as
\begin{equation}\label{Ent3}
dS_{\sum}=-\frac{3\zeta(3)\gamma \sqrt{-r_v ^2f(r_v)}}{2} \left(\frac{\beta}{ A }\right)dm,
\end{equation}
The first law of black hole thermodynamics for spherically symmetric rotating black holes is
\begin{equation}\label{1stlaw}
{dm}=\frac{d{S}_{BH}}{\beta}+\Omega_{H} dJ,
\end{equation}
here $S_{BH}$ is the Bekenstein-Hawking entropy of black hole. As an angular momentum is a conserved quantity hence its distortion will be small at the horizon and will be negligibly small at $r=r_v=0.45m$. Thus, we will not consider the effect of angular momentum in the onward discussion. The above Eq. (\ref{1stlaw}) becomes
\begin{equation}
{dm}=\frac{d{S}_{BH}}{\beta},
\end{equation}

\begin{figure}
\begin{center}
\includegraphics[width=0.6\textwidth]{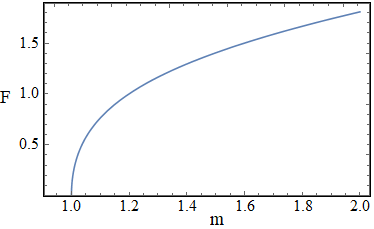}
\caption{Plot of evolution relation $F(m )$ vs. mass $m$ for rotating BTZ black hole.}
\label{image-3}
\end{center}
\end{figure}

Using this simplified form of first law black hole thermodynamics in Eq. (\ref{Ent3}), we get a relation between the interior and exterior entropy of black hole as
\begin{equation}
dS_{\sum}=-\frac{3\zeta(3)\gamma \sqrt{-r_v ^2f(r_v)}}{2} \left(\frac{dS_{BH}}{ A }\right),
\end{equation}
This equation gives a direct relationship between the two types of entropy. Since, we know the quantum modes entropy is directly related to the interior volume of the black hole, it could be maximized by the polynomial $\sqrt{-r_v ^2f(r_v)}$. So, we call this maximal interior entropy. On the other hand, the area $A$ of the black hole will be constant at $r=r_v$. Both of them are mass-dependent. This means that the relation between the interior and exterior entropy of a black hole is a function of mass $m$. Generally, this connection between the two types of entropy can also be expressed as
\begin{equation}\label{proprela}
 dS_{\sum}=-\frac{4 \gamma \zeta (3)}{3 \pi }F(m)d{S}_{\text{BH}},
\end{equation}
here $F(m)$ is the evolution function and is given by
\begin{equation}\label{Proprel}
F(m)=\frac{\sqrt{m \left(3 X^2+\sqrt{3 X^2+1}-1\right)}}{\sqrt{X+1}},
\end{equation}
The negative sign in Eq. (\ref{proprela}) shows the quantum modes entropy increases with Eddington time whereas, Bekenstein-Hawking entropy decreases. The curve in Fig. (\ref{image-3}) shows the evolution relation as a function of black hole mass. This plot is similar to the power function of some variable having power as a fraction between $0$ and $1$. This plot shows that as the BTZ black hole mass $(m)$ increases from zero, the slope of the curve also increases. In the beginning, the BTZ black hole mass is seen to be constant for some increase in evolution relation or slowly increases and gaining some mass limit the evolution relation increases with an increase in black hole mass continuously without any divergence.

\section{Remarks and discussion}\label{Sec.4}

We analyzed $(2+1)-$ dimensional uniformly rotating BTZ black hole with conserved angular momentum $J$ for understanding its interiors informations. An imaginary hyper-surface extending from the event horizon to the center of the black hole is impinged as shown in Fig. (\ref{image-1}. The part of hyper-surface that contributes to the increase of black hole's interior volume is calculated and found that the maximal volume bounded by a black hole could be obtained by maximization of the factor $\sqrt{-r^2 _vf(r_v)}$ as given in Eq (\ref{maxhyp}). From this equation, it is verified that the interior volume of a rotating BTZ black hole is linearly increasing with $v$. Numerically, its position of hyper-surface found to be $r_v=0.45m$ given in Fig. (\ref{image-2}). It is the most valuable task in defining an interior volume. So, one needs to be careful at this stage.

The quantum mode entropy is a linear function of black hole interior volume and hence of Eddington time. So, the relation of $v$ with black hole interior volume and quantum modes entropy is considered a game changer for statistical quantities in the interior of the black hole. Thinking so, we supposed two assumptions for confirmation of changes in statistical quantities with $v$. The first of these assumptions guaranteed us to use Boltzmann law and the second one led us to investigate the differential form of interior quantum modes entropy. This differential form of quantum mode entropy gives an evolution relation with the differential form of Bekenstein-Hawking entropy as a connection between interior and exterior entropy of BTZ black hole. According to Parikh \cite{Parikh:2005qs}, this evolution relation between the interior and exterior entropy will enable us to solve the problem of black hole information paradox. The results obtained from this discussion will be in-sighted in the next section (\ref{Sec.5}).

An important point to be noted here is the mechanism of Hawking radiation emission and its effect on the interior volume and entropy. In our investigations, the interior volume and quantum mode entropy are verified to increase with Eddington time. In terms of Hawking temperature and entropy Eq. (\ref{BTZvoleq}) and (\ref{Ent1}) with Eq. (\ref{Hawent}) and (\ref{Temp}) can be written as
\begin{equation}
V_{CR}=\frac{\pi S_{BH}T_H}{l}v, \qquad S=\frac{3\zeta(3)S_{BH}T^3}{8l}v,
\end{equation}
It shows 
\begin{equation}
\partial_v V_{CR}>0, \qquad \partial_v S>0 \qquad \partial_v m<0,
\end{equation}
or we can also write 
\begin{equation}
\partial_m V_{CR}<0, \qquad \partial_mS<0,
\end{equation}
From these inequalities, if the black hole losses its mass and angular momentum during the Hawking radiation then it must recover the Hawking temperature \cite{Sakalli:2015, Zou:2014gla}. It means that the tunneling rate is independent of particle type and the emission process must be slow enough to leave the horizon entropy and Hawking temperature nearly unchanged. Based on this point, we assumed the emission process is quasi-static. Similarly, for a radiating black hole, the probability of states is given by $e^{S_{BH}}$. It shows that while reduction of the black hole, only a few quantum states will be available for entanglement with Hawking radiation. Alternatively, there will be more quantum states to store the information. It is also a confirmation of our results.
\section{Conclusions}\label{Sec.5}
Black hole's interior volume is a dynamical and time-dependent quantity. We used Baocheng Zhang \cite{Zhang:2015gda} mechanism to investigate interior volume bounded by the rotating BTZ black hole. We found the maximal interior volume associated with the largest hyper-surface linearly increasing with $v$. The position of hyper-surface is numerically found at $r_v=0.45m$ as shown in Fig. (\ref{image-2}). This figure shows that all hyper-surfaces associated with other points bounded by rotating BTZ black hole are smaller than $r_v=0.45m$. So, $r_v$ will bound the maximal interior volume. In comparison to other types of rotating black holes (like Kerr and Kerr Newman black holes), these results are different and interesting due to space-time geometry in the interior of the BTZ black hole.

As there is a direct relationship between the interior volume and quantum modes entropy hence quantum modes entropy is also proportional to $v$. It means that the relation between the quantum modes entropy and $v$ can affect the statistical quantities in the interior of the black hole. To proceed for some result in agreement with this statement, we calculated the total number of quantum modes (\ref{Qmodes}), free energy (\ref{FE}) and the entropy (\ref{Ent1}) of the scalar field in the interior of the black hole at some inverse temperature $\beta$.

To study the effects of linear relation between quantum mode entropy and $v$ on statistical quantities in the interior of black hole, we considered two important assumptions that led us to use Boltzmann law and write the differential form of quantum mode entropy. Using differential form, an evolution relation is obtained between quantum mode entropy and Bekenstein-Hawking entropy under Hawking radiation. A surprising feature of this evolution relation $F(m)$ is its dependence on black hole mass $m$ as given in Eq. (\ref{Proprel}). We have a qualitative equation for this as $dS=-(am)dS_{BH}$ with $a$ as a constant.

The behavior of this result in comparison with a rotating black hole (like Kerr's black hole) is quite different see Fig (\ref{image-3}). Recall Fig (3) in Ref. \cite{Wang:2018dvo}, when the mass of Kerr black hole goes to infinity, the curve of proportionality relation approaches to a constant value. While in the case of rotating BTZ black hole, the curve is divergent as shown in Fig. (\ref{image-3}). It means that there appears some important phenomena ought to understand for this difference. The lower portion of Fig. (\ref{image-3}) is for a relatively small mass where the Stephan Boltzmann law is valid only when $m\gg J$, but here this condition of Hawking radiation will be hardly satisfied because it will approach an extreme state. So, the above-investigated result between the two types of entropy seems impractical at this position.

The Physical significance of this work is the extension of interior volume and entropy notion to the lower space-time dimensions. For relatively large BTZ black holes, the proportionality function is a linear function of the black hole's mass while its plot is divergent as compared to other black holes. It gives a new background to understand the solution of information paradox issue in lower-dimensions.
\section*{Acknowledgments}
We are extremely thankful to anonymous reviewers for their valuable comments and suggestions that help us to improve our manuscript. This work is supported by National Academy of Higher Education, High Education Commission of Pakistan (Grant No. 17/IPFP-II(Batch-I)/SRGP/NAHE/HEC/2020/114).

\end{document}